\documentclass[preprint2]{aastex}
\usepackage{spr-astr-addons}
\usepackage{graphicx}
\usepackage{latexsym}
\usepackage{amssymb}
\usepackage{amsmath}

\newcommand{\vrr}{v_{r}}
\newcommand{\vf}{v_{\phi}}            
\newcommand{\vk}{v_{K}}               
\newcommand{\vz}{v_{z}}               
\newcommand{\ee}[1]{\times 10^{#1}}

\begin{document}
\title{Radial and vertical angular momentum transport in protostellar discs}

\shorttitle{Angular momentum transport in protostellar discs}        
\shortauthors{Salmeron et al.}

\author{Raquel Salmeron \altaffilmark{1}}
\affil{Research School of Astronomy and Astrophysics and Research School of Earth Sciences, The 
Australian National Universty. }
\author{Arieh K\"onigl}
\affil{Department of Astronomy \& Astrophysics, The University of Chicago} 
\author{Mark Wardle}
\affil{Physics Department, Macquarie University}

\altaffiltext{1}{Previous address: Department of Astronomy \& Astrophysics, The University of Chicago.}



\begin{abstract}
Angular momentum  in protostellar discs can be transported either
radially, through turbulence induced by the magnetorotational
instability (MRI), or vertically, through the torque exerted by a
large-scale magnetic field. We present a model of steady-state discs where these two mechanisms operate at the same radius and derive approximate criteria for their occurrence in an ambipolar diffusion dominated disc.  We obtain
``weak field''
solutions -- which we associate with the MRI channel modes in a stratified disc -- 
and transform them into accretion solutions with
predominantly radial angular-momentum transport by implementing a
turbulent-stress prescription based on published results of numerical simulations. We also analyze ``intermediate field strength'' solutions
in which both radial and vertical transport operate at the same radial
location. Our results suggest, however, that this overlap is unlikely to occur in real discs.
\end{abstract}


\section{Introduction}
\label{intro}

Angular momentum in protostellar discs can be transported either radially, through turbulent viscosity induced by the magnetorotational instability (MRI; e.g. Balbus \& Hawley 1998); or vertically, by centrifugally driven winds (e.g. K\"onigl \& Pudritz 2000).  It is likely, in fact, that both forms of transport
play a role in real discs.\footnote{\label{note} Note that vertical angular-momentum transport could also 
occur via other mechanisms, not considered here; such us magnetic braking
(e.g. Krasnopolsky \& K\"onigl 2002), ``failed'' winds or non-steady phenomena.} 
Quasi-steady disc models that incorporate the two mechanisms have already been attempted (e.g. Lovelace et al. 1994; Casse \& Ferreira
2000; Ogilvie \& Livio 2001), but these studies treated the radial transport using prescriptions that do not account for its origin in MRI-induced turbulence.  A more refined approach seems to be needed, however, as solutions in which a wind carries away all the excess angular momentum of the fluid are not susceptible to sustain MRI-unstable modes (Wardle \& K\"onigl 1993, hereafter WK93).
 Our prescription to incorporate radial transport in our wind-driving models uses the fact that 
the growth (and structure) of the MRI is dependent on the local strength of the magnetic field, which we measure by $a$, the ratio of the Alfv\'en speed to the isothermal
sound speed (see section \ref{sec:vertical}). MRI modes are  expected to grow
when $a \ll 1$ (provided that the gas is sufficiently coupled to the field), but  they are suppressed when $a$ is comparatively large ($\gtrsim 1$). Since $a$ ($\propto \rho^{-0.5}$) increases with height,
this suggest that radial and vertical angular momentum transport could in principle operate at the same radius, with the MRI acting in a finite section next to the midplane, and outflows dominating at higher $z$.

Since protostellar discs are weakly ionized over most of their extent, it is essential to incorporate the effect of finite magnetic diffusivity in our models. This is done in our general scheme by employing a 
tensor-conductivity, obtained via a realistic vertical ionization structure. This formulation naturally
incorporates the three relevant magnetic diffusion mechanisms (i.e.~ambipolar, Hall and Ohmic). The illustrative examples presented here, however, assume that ambipolar diffusion dominates over the entire section of the disc and (except in section \ref{sec:combined}), take the magnetic coupling to be constant with height. This is useful in order to apply the analytic results obtained by  WK93 for this case to our prescription for MRI-induced turbulent transport in a wind-driving disc model. We refer the reader to Salmeron, K\"onigl \& Wardle 2007 (hereafter  SKW07) for full details of the study. Section \ref{sec:vertical} presents the wind-driving model. Section \ref{sec:radial} outlines our prescription for MRI-induced radial transport and illustrates it with a solution with a predominantly radial angular-momentum transport. Section \ref{sec:combined} discusses a representative solution that incorporates radial and vertical transport and our findings are summarised in section \ref{sec:conclusion}.

\section{Centrifugally driven winds}
\label{sec:vertical}

Bipolar outflows are commonly associated with protostellar accretion discs, as well as other astronomical accreting sources. Their ubiquity suggests that they play a key role in regulating accretion, a notion supported by the apparent correlation between accretion and outflow signatures in these systems (e.g. Hartigan, Edwards \& Ghandour 1995).
It is also generally accepted that these outflows are highly collimated winds, accelerated centrifugally from the surfaces of discs by the mechanism 
first proposed by Blandford \& Payne (1982, hereafter BP82). According to this picture, disc material is flung out if the open magnetic field lines that thread the disc are sufficiently inclined (i.e. the angle between the line and the disc plane is $< 60\,^{\circ} $), initiating an outflow that can potentially reach super Alfv\'enic speeds. 

A key feature of these winds, and one of particular interest in the context of this work, is that they provide an efficient means of extracting the rotational kinetic energy of the disc and carrying away the excess angular momentum of the accreted matter. 
This is expected to be the dominant angular momentum transport mechanism in discs when the magnetic field is strong (suprathermal) and the MRI is suppressed because the wavelength of the most unstable perturbations exceed the magnetically reduced disc scale height. Wind-driving solutions from strongly magnetised discs have been obtained semi-analytically and numerically (e.g.~WK93; Li 1995, 1996; Casse \& Keppens 2002).  

In this work we have used the modelling procedure developed by WK93 to obtain radially localised disc-wind solutions from strongly magnetised discs. This entails solving the conservation (of mass and momentum) and induction equations, as well as Amp\`ere's and Ohm's laws for a steady-state, isothermal,
geometrically thin and nearly Keplerian accretion disc. We neglect
all radial derivatives except that of $v_\phi$ and use a conductivity tensor (e.g. Wardle 1999) to communicate the effect of the charged species on the neutrals.  

These equations are integrated vertically upwards to obtain the disc solution and the location of the critical (sonic) point. This solution is then matched onto a BP82, self-similar wind solution, by imposing the regularity condition at the Alfv\'en critical point. The relevant model parameters are
(1)  $a_0 = v_{{\rm A}0}/c_{\rm s}$, the midplane (subscript 0) ratio of the Alfv\'en
speed to the sound speed, which measures the magnetic field strength; 
(2) $\eta$, the ratio of the Keplerian rotation time to the neutral--ion momentum-exchange time, which determines the magnetic coupling; 
(3) $\epsilon \equiv -v_{r0}/c_{\rm s}$, the midplane (normalized) inward radial
speed, evaluated by imposing the Alfv\'en critical-point constraint on the wind;
(4) $c_{\rm s}/v_{\rm K} = h_{\rm T}/r$, the ratio of the disc tidal scale
height to the radius; and  
(5) $\epsilon_B \equiv -cE_{\phi}/c_{\rm s} B_z$, the normalized azimuthal
component of the electric field, which is nonzero if the magnetic field lines drift
radially (WK93).\footnote{\label{ansatz}Setting 
$\epsilon_{\rm B} = 0$ effectively fixes the value of $B_r$ at the disc surface (subscript `b'). In a 
global formulation $B_r$ would be determined by the radial distribution of $B_z$ 
(e.g. Ogilvie \& Livio 2001; Krasnopolsky \& K\"onigl 2002). }

The main properties of
wind-driving, strongly magnetized ($a_0 \lesssim 1$) discs are analysed in  WK93. Fig.~1 of SKW07 shows a solution of this type that matches
on to a global BP82 wind solution (see also the pure-wind case of Fig.~\ref{fig:mri}). These solutions satisfy:
\begin{equation}
(2\eta)^{-1/2} \lesssim a_0 \lesssim \sqrt{3}  \lesssim \epsilon\eta \, \,;  \qquad \eta > 1 \,.
\label{eq:ineq}
\end{equation}
The first inequality in (\ref{eq:ineq}) expresses the
requirement that $v_\phi < v_{\rm K}$ in the disc region, the second gives the wind launching condition
($B_{r,b}/B_z > 1/\sqrt{3}$, where subscript `b' denotes the disc surface), the third 
constrains the base of the wind to lie above the magnetically-reduced density scale height ($h$) and
the fourth is the minimum field-matter coupling condition. 

\section{MRI-induced turbulence}
\label{sec:radial}
The MRI transports angular momentum radially outwards as fluid elements exchange angular momentum non-locally via the distortion of the magnetic field lines that connect them. 
Its properties have been extensively investigated in both the linear and nonlinear stages 
(e.g. Sano \& Stone 2002a,b; Salmeron \& Wardle 2003, 2005).
In particular, numerical simulations show that the MRI sustains radial angular-momentum
transport when the initial (subscript i)
Elsasser number $\Lambda_{\rm i} \equiv v_{{\rm A0},{\rm
i}}^{2}/\eta_{\rm Ohm}\Omega_{\rm K}$ is $\gtrsim 1$ (e.g. Sano \&
Inutsuka 2001; Sano \& Stone 2002a,b), where $\eta_{\rm Ohm}$ is the Ohmic
diffusivity.  For the vertical initial field geometry
assumed here, it turns out that $\Lambda_{\rm i} = \eta_{\rm i}$ (SKW07) and the minimum-coupling condition is the same for both forms of angular-momentum
transport. So the question we seek to answer is: Assuming that $\eta \gtrsim 1$, which section of the disc (if any) would be unstable to the MRI?
\begin{figure}[ht!]
\begin{center}
\includegraphics[width=0.45\textwidth]{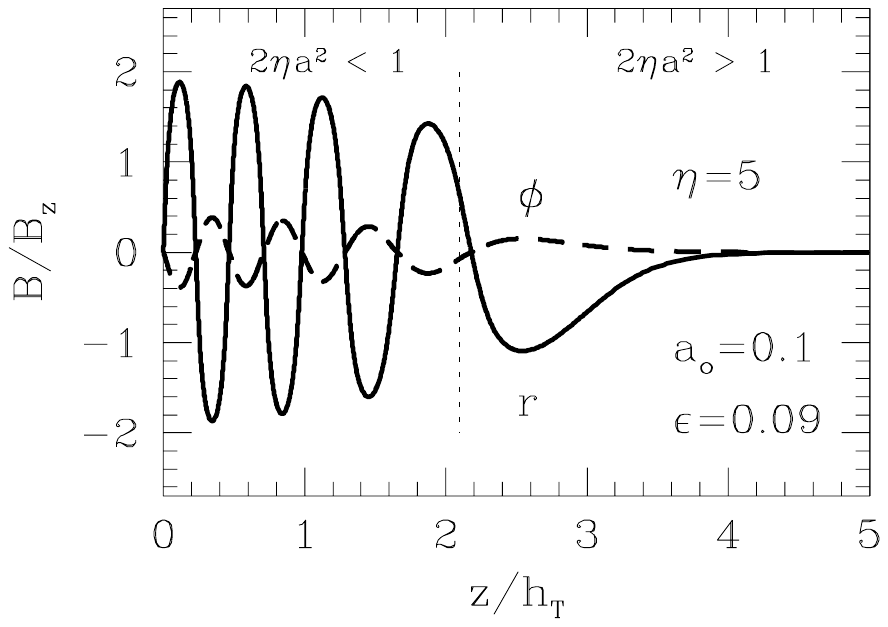}
\end{center}
\caption{Structure of a weakly magnetized disc with vertical magnetic angular-momentum transport in the limit where no wind develops. The radial and vertical components of the magnetic field are indicated by the thick solid and dashed lines, respectively. The vertical dotted line marks the boundary of the MRI-unstable zone (see text).}
\label{fig:wiggles}
\end{figure}
The first inequality in (\ref{eq:ineq}) can in fact be used to derive a
lower limit of $a$ for MRI stability. This is the case because when this expression is
violated, the surface layers of the disc become super-Keplerian and experience 
outward streaming motion that is unphysical for a pure
wind solution (WK93), but is \emph{characteristic} of the two-channel
mode that underlies the MRI. Moreover, wind-driving disc solutions that satisfy all the constraints in (\ref{eq:ineq}) are {\em stable} to the fastest growing linear mode of the MRI (WK93), a condition that further supports this choice. The above constraint on $a$ was applied by WK93 at
the midplane, since in their solutions the wind carries away 
the entire excess angular momentum of the gas. Here, however, it is more appropriate to apply it {\em
locally} [using $a(z) \propto \rho(z)^{-1/2}$]  to differentiate
 the MRI-unstable section of the disc (where
$2\eta a^2<1$) from the region where only vertical angular
momentum transport takes place ($2\eta a^2>1$).

Fig.~\ref{fig:wiggles} shows a solution where $a<(2\eta)^{-1/2}$ over most 
of the disc vertical extent. Note that no wind develops and thus, angular momentum 
is transported magnetically between different heights and ultimately 
carried away radially by the flow.  As anticipated, both the radial and azimuthal field components oscillate in the region where $2\eta a^2
\lesssim 1$ and the disc sections where $B_z \partial B_\phi/\partial z < 0$ \ $(> 0)$
lose (gain) angular momentum and have $v_r <0$ \ $ (> 0)$. We associate this solution with a two-channel MRI mode for a stratified, non-ideal MHD disc. These modes were analysed by  
Goodman \& Xu (1994) in the unstratified,
ideal-MHD case. These authors showed that they are exact solutions in the linear and nonlinear regimes, but are {\em unstable} to parasitic modes. Although a rigorous stability analysis of the modes in stratified, non-ideal MHD discs have not yet been done, numerical simulations suggest that they may indeed evolve into turbulence (e.g.~Hawley et al.~1995;
Sano et al.~2004, hereafter S04). 

Having derived a criterion to identify the MRI-unstable section of the disc ($\eta\gtrsim 1$ and $2\eta a^2\lesssim 1$), we now need to quantify the turbulent angular momentum transport that we expect
to develop in these regions. To accomplish this, we use the results of numerical simulations in the literature to
evaluate the spatial and temporal average of the $r \phi$ component of the turbulent Maxwell stress (the main component of the stress tensor) in terms of the
similarly averaged magnetic pressure, obtaining
 \begin{equation}
\ll w_{r\phi} \gg\,  \approx 0.5 \, \frac{\ll B^2 \gg}{8 \pi}
\label{eq:wrphi}
\end{equation}  
(e.g. Hawley et al.~1995; S04). Using the characteristic ratios
for the field components in MRI turbulence reported in S04, this can be expressed as
\begin{equation}
\label{eq:mri}
\ll w_{r\phi} \gg \approx 14\,  Y\,  \frac{B_{z,{\rm i}}^2}{8 \pi}\ ,
\end{equation}
where $Y \equiv \, \ll B_z^2 \gg /B_{z,{\rm i}}^2 $ is also listed in 
S04 as a function of $\beta_{\rm i} \equiv2/a_{\rm i}^2$, the 
initial plasma $\beta$ parameter. Finally, the radial angular-momentum transport associated with this term is added to the  angular momentum conservation equation in the disc region deemed to be unstable to the MRI. This equation then reads
\begin{equation} 
\label{eq:momentum} 
\frac{\rho \vrr \vk}{2r} + \rho \vz
\frac{d \vf}{dz} = -\frac{J_r B_z}{c} - \frac{\ll w_{r\phi}\gg}{r}\ ,
\end{equation} 
where we used $(1/r)\partial (r^2 \ll w_{r\phi} \gg )/ \partial r \approx \, \ll  w_{r\phi}\gg$,  $\partial(r \vf)/\partial r \approx \vk/2$ and $J_z
\approx 0$. Note that we are implicitly assuming that the turbulent magnetic field
components only contribute to the radial angular-momentum transport and
interpret the other magnetic-field (and velocity) terms in the disc structure equations as the {\em mean} values of the respective quantities.  Similarly, we take the first
term on the r.h.s. of equation~(\ref{eq:momentum}) as the vertical angular-momentum transport induced by the mean field. The underlying assumptions of this approach
 would need to be verified numerically.

\section{Combined angular momentum transport}
\label{sec:combined}

We now explore solutions where both radial and vertical angular momentum transport take place at the same radial location in the disc.  From the above discussion, it is clear that they 
 require moderate coupling ($\eta \sim 1$) and field strengths [$a_0 < (2\eta_0)^{-1/2}$ but $a(z=z_{\rm b}) \sim 1$] for a wind to develop
\emph{and} the MRI to be active over a sizable section of the disc. To satisfy these requirements, we have allowed $\eta$ to decrease with $z$ and fall slightly below 1 in these solutions (with the expectation that it is still large enough to sustain MRI turbulence; see Sano \& Stone 2002a,b). The adopted profile (with $\eta$ decreasing from $0.63$ to $0.5$ between the midplane and the surface) is motivated by the inferred behavior of $\eta(z)$ in the ambipolar-diffusion dominated regions of protostellar disks (see SW07 and Salmeron \& Wardle 2005).
\begin{figure*}[ht!]
\begin{center}
\includegraphics[width=0.8\textwidth]{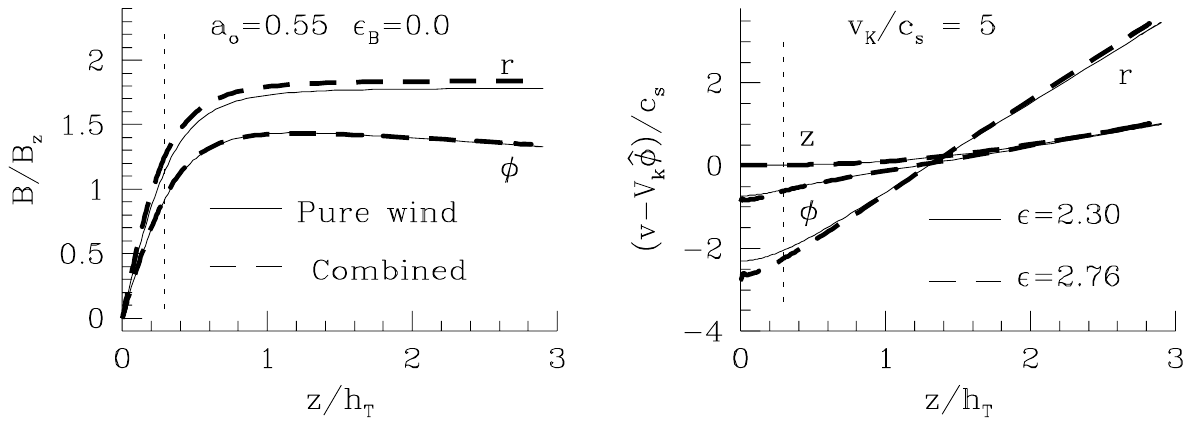}
\end{center}
\caption{Wind-driving disc with a moderately strong field and 
$\eta$ decreasing from $0.63$ to $0.5$ between $z = 0$ and $z = z_b$. 
Density, magnetic field components ({\it left}) and velocity components 
({\it right}) are shown for
a pure wind 
solution (solid lines) and also for a case where radial transport is included (dashed lines) in the region where $2\eta a^2 < 1$, or to the left of the vertical dotted lines. The curves terminate at the respective sonic points.}  
\label{fig:mri} 
\end{figure*}

Fig.~\ref{fig:mri} shows 
solutions obtained under these assumptions.  The solid lines display a pure wind (vertical angular-momentum transport only) solution. The dashed lines depict how this solution is modified when radial angular-momentum transport is incorporated via the prescription introduced in Section 3. 
 Both solutions are compared in Table~\ref{table:boundary}. Note that the addition of radial angular-momentum transport results in this case
in an increased inflow speed (measured by $\epsilon$) and a higher mass
inflow rate ($\dot{M}_{\rm in}$).  The higher $|v_r|$ corresponds
to a stronger radial neutral--ion drag (for $\epsilon_{\rm B}=0$;
see footnote~\ref{ansatz}), and therefore to a stronger radial bending of the magnetic
field lines (e.g.~a higher $B_{r,{\rm b}}$). This, in turn, increases the level of magnetic
squeezing (reflected in the lower value of $h$),  and leads to a stronger
density stratification (e.g.~a lower $\rho_{\rm s}/\rho_0$). Although the disc thickness 
does not change appreciably ($z_{\rm b} = 1.25\, h_{\rm T}$ in both cases) and the height
of the sonic point ($z_{\rm s}/h_{\rm T}$) decreases (from $2.90$ to $2.82$), the lower $\rho_{\rm s}/\rho_0$ leads to a lower wind outflow
rate and vertical torque (measured by $\kappa$ and $T_z$, respectively) in the combined solution.

The above results  suggest that the joint operation of
these two angular-momentum transport mechanisms might be self-limiting (in which case it would be unlikely to introduce a dynamical instability): The incorporation of MRI-induced
turbulence increases the overall angular momentum transport in the disc, but {\em reduce} the amount of
angular momentum carried away by the wind. This conclusion could be tested through an explicit
stability analysis (e.g. K\"onigl 2004) or via numerical simulations.

\begin{table}[ht!]
\caption{Comparison of the solutions in Fig.~\ref{fig:mri}. 
$\dot M_{\rm in}$ is the mass accretion rate per disc circumference,
normalized by $(\rho_0 c_{\rm s} h_{\rm T})$.  The vertical ($T_z$)  and radial ($T_r$)
torques per unit area are normalized by $r_0 B_0^2/4\pi$. All heights are listed in units of
$h \equiv z(\rho = \rho_0/\sqrt{e}$).} 
\begin{tabular}{|c|c|c|}
\hline
Disc/outflow & Pure wind & Combined\\
properties  & solution & solution \\
\hline
$h/h_{\rm T}$ & $0.31$ & $0.29$ \\
$z_{\rm b}/h$ & $3.99$ & $4.35$ \\
$z_{\rm s}/h$ & $9.25$ & $9.81$ \\    
$\rho_s/\rho_0$ & $1.0\ee{-2} $ & $8.3\ee{-3} $ \\
$\kappa$ & $0.17$ & $0.14 $ \\
$B_{r,{\rm b}}/B_z$ & $1.75$ & $1.81$ \\
$|B_{\phi,b}|/B_z$ & $1.44$ & $1.43$ \\
$T_z$ & $1.63$ & $1.59$ \\
$T_r$ & --- & $0.09$ \\
$\dot M_{\rm in}$  & $1.74$ & $1.84$ \\   
\hline                                
\end{tabular}
	\label{table:boundary}
\end{table}

\section{Discussion}
\label{sec:conclusion}

We have examined the two leading processes thought to be responsible for transporting angular momentum in protostellar discs: Jets and outflows accelerated centrifugally from the disc surfaces and turbulent viscosity induced by the magnetorotational instability.  Our main motivation has been twofold. First, we have set out to explore these processes under more realistic conditions, taking into account the detailed ionisation structure and vertical stratification of the disc. Second, we are interested in the possibility that both forms of transport could operate at the same radial location and, if so, what is their relative importance at different locations and evolutionary stages in protostellar discs. 
Our overall formulation incorporates a height dependent, tensor conductivity. This enables us to model the different diffusion mechanisms available to the fluid (ambipolar, hall and ohmic) and how their relative contributions change with location. The illustrative examples presented here, however, assume ambipolar diffusion dominates over the entire section of the disc and the magnetic coupling is constant with height. Our main results are highlighted below.
\begin{enumerate}
\item We derived an approximate criterion to identify the section of the disc that would be unstable to MRI-induced turbulence and developed a scheme, based on the results of published numerical simulations, to quantify the amount of angular momentum transported radially in this region. 
\item  We showed a weak field solution exhibiting
oscillatory field and velocity profiles and associated it with the nonlinear MRI channel mode
identified by Goodman
\& Xu (1994) in an ideal-MHD, unstratified case.  A rigorous stability analysis of our solutions has not been attempted yet, but evidence from linear analyses and
numerical simulations suggests that they may be unstable. 
\item We argued that radial angular-momentum transport operates 
where $2 \eta a^2 < 1$ and presented a hybrid solution in
which this transition occurs at an intermediate height below the launching region. 
\item This solution shows that the addition of radial transport  increases the total amount of angular momentum removed
from the inflowing gas, but reduces the angular momentum flux carried away by the wind.
\item It appears that these two transport mechanisms are unlikely to 
have a significant spatial  overlap in real discs, as this requires a moderate $a_0$ ($\gtrsim 0.5$) and low (and decreasing with height) values of $\eta$
($\eta_0\lesssim 1$) that are unlikely to be attained over measurable radial extents. 
\end{enumerate}

We are currently in
the process of applying our general scheme to the investigation of 
various conductivity regimes in realistic disc models. This would enable us to estimate the fractions of angular momentum transported via turbulent viscosity and outflows, respectively; and assess the implications of the resulting disc structure to other discs processes, in particular to planet formation and migration.

\section*{Acknowledgments}

We thank the referee for useful comments. This research was supported by NASA Theoretical Astrophysics Program 
grant NNG04G178G and the Australian Research Council.


\begin{thebibliography}{}

\bibitem[]{} 
Balbus S.A., Hawley J.F., 1998, Rev. Mod. Phys., 70, 1 


\bibitem[]{} 
Blandford R.D., Payne D.G., 1982, MNRAS, 199, 883 (BP82) 

\bibitem[]{} 
Casse F., Ferreira J., 2000, A\&A, 353, 1115 

\bibitem[]{} 
Casse F., Keppens R., 2002, ApJ, 581, 988

\bibitem[]{}
Goodman J., Xu G., 1994, ApJ, 432, 213

\bibitem[]{}
Hartigan P., Edwards S. Ghandour L., 1995, ApJ, 452, 736 

\bibitem[]{} 
Hawley J., Gammie C., Balbus S., 1995, ApJ, 440, 742 

\bibitem[]{}
K\"onigl A., 1989, ApJ, 342, 208 

\bibitem[]{}
K\"onigl A., 2004, ApJ, 617, 1267 

\bibitem[]{}
K\"onigl A., Pudritz R. E., 2000, in Mannings, V. G., Boss, A. P.,
Russell, S. eds, Protostars \& Planets IV. Univ. Arizona Press,
Tucson, p. 759

\bibitem[]{}
Krasnopolsky R., K\"onigl A., 2002, ApJ, 580, 987

 
\bibitem[]{}
Li Z.-Y. 1995, ApJ, 444, 848

\bibitem[]{}
Li Z.-Y. 1996, ApJ, 465, 855

\bibitem[]{}
Lovelace R. V. E., Romanova M. M., \& Newman W. I., 1994, ApJ, 437, 136
 
\bibitem[]{}
Ogilvie G. I., Livio M., 2001, ApJ, 553,158

\bibitem[]{}
Salmeron R., K\"onigl A., Wardle M., 2007, MNRAS, 375, 177 (SKW07)

\bibitem[]{}
Salmeron R., Wardle M., 2003, MNRAS, 345, 992 

\bibitem[]{}
Salmeron R., Wardle M., 2005, MNRAS, 361, 45

\bibitem[]{}
Sano T., Inutsuka S., 2001, ApJ, 561, L179

\bibitem[]{}
Sano T., Stone J. M., 2002a, ApJ, 570, 314

\bibitem[]{}
Sano T., Stone J. M., 2002b, ApJ, 577, 534

\bibitem[]{}
Sano T., Inutsuka S., Turner N., Stone J. M., 2004, ApJ, 605, 321 (S04)

\bibitem[]{}
Wardle M., 1999, MNRAS, 307, 849

\bibitem[]{}
Wardle, M., K\"onigl, A. 1993, ApJ, 410, 218 (WK93) 

\end{thebibliography}
\end{document}